\documentclass[a4paper,11pt]{article}
\pdfoutput=1 

\usepackage{jcappub} 

\usepackage[T1]{fontenc} 
\usepackage{multirow}

\usepackage{ulem}   % a package related to strikethrough
\usepackage{xcolor}
\usepackage{bm}

\definecolor{grey}{rgb}{0.1,0.1,0.6}

\title{\boldmath Quasinormal modes of black holes in $f(T)$ gravity}

\author[a,b]{Yaqi Zhao}
\author[a,b]{Xin Ren}
\author[a,b]{Amara Ilyas}
\author[c,a,b,1]{Emmanuel N. Saridakis}
\author[a,b,1]{Yi-Fu Cai,\note{Corresponding author.}}

\affiliation[a]{Deep Space Exploration Laboratory/School of Physical Sciences, University of Science and Technology of China, Hefei, Anhui 230026, China}
\affiliation[b]{CAS Key Laboratory for Researches in Galaxies and Cosmology/Department of Astronomy, School of Astronomy and Space Science, University of Science and Technology of China, Hefei, Anhui 230026, China}
\affiliation[c]{National Observatory of Athens, Lofos Nymfon, 11852 Athens, 
Greece}

\emailAdd{zxmyg86400@mail.ustc.edu.cn}
\emailAdd{rx76@mail.ustc.edu.cn}
\emailAdd{aarks@ustc.edu.cn}
\emailAdd{msaridak@phys.uoa.gr}
\emailAdd{yifucai@ustc.edu.cn}

\abstract{We calculate the quasinormal modes (QNM) frequencies of a test massless scalar field and an electromagnetic field around static black holes in $f(T)$ gravity. Focusing on quadratic $f(T)$ modifications, which is a good approximation for every realistic $f(T)$ theory, we first extract the spherically symmetric solutions using the perturbative method, imposing two ans$\ddot{\text{a}}$tze for the metric functions, which suitably quantify the deviation from the Schwarzschild solution. Moreover, we extract the effective potential, and then calculate the QNM frequency of the obtained solutions. Firstly, we numerically solve the Schr$\ddot{\text{o}}$dinger-like equation using the discretization method, and we extract the frequency and the time evolution of the dominant mode applying the function fit method. Secondly, we perform a semi-analytical calculation by applying the WKB method with the Pade approximation. We show that the results for $f(T)$ gravity are different compared to General Relativity, and in particular we obtain a different slope and period of the field decay behavior for different model parameter values. Hence, under the light of gravitational-wave observations of increasing accuracy from binary systems, the whole analysis could be used as an additional tool to test General Relativity and examine whether torsional gravitational modifications are possible.}

\begin{document} 
\maketitle
\flushbottom

\section{Introduction}
\label{sec: introduction}

Modified gravity is one possible direction that we can follow to phenomenologically describe two phases of universe acceleration, at early and late times respectively \cite{Starobinsky:1980te, CANTATA:2021ktz, Capozziello:2011et}, while possessing the additional advantage of being closer to a quantum description of gravity \cite{Addazi:2021xuf}. The simplest way to construct modified theories of gravity is to start from the standard curvature-based formulation, namely General Relativity (GR), and extend it in various ways \cite{Capozziello:2002rd, Nojiri:2010wj, Nojiri:2005jg, Nicolis:2008in, Clifton:2011jh}. Alternatively, one can start from the teleparallel equivalent of general relativity (TEGR) \cite{Aldrovandi:2013wha, Krssak:2018ywd}, and arrive at a series of torsion-based modified theories of gravity \cite{Cai:2015emx, Bengochea:2008gz, Linder:2010py, Chen:2010va, Bahamonde:2015zma,  BeltranJimenez:2018vdo, Li:2018ixg, Bahamonde:2019shr, Bahamonde:2021gfp}.

Although TEGR is equivalent to GR at the level of field equations, their nonlinear extensions are not equivalent, resulting in different classes of modification. The simplest nonlinear torsional modified gravity is $f(T)$ gravity, which has the advantage that its  field equations have up to second order derivatives. Its cosmological applications prove to be very interesting 
\cite{Ferraro:2006jd, Cai:2019bdh, Yan:2019gbw, Wang:2020zfv, Bose:2020xdz, Ren:2021tfi, Escamilla-Rivera:2021xql} and the theory has been shown to be in agreement with observations \cite{Wu:2010mn, Cardone:2012xq, Nesseris:2013jea, Nunes:2016plz, Basilakos:2018arq, Xu:2018npu, Ren:2022aeo}. Furthermore, the black holes solutions in $f(T)$ gravity have been investigated in detail \cite{Boehmer:2011gw, Gonzalez:2011dr, Meng:2011ne, Ferraro:2011ks, Pfeifer:2021njm, Wang:2011xf, Atazadeh:2012am, Rodrigues:2013ifa, Nashed:2013bfa, Bejarano:2014bca, Nashed:2014iua, Junior:2015fya, Das:2015gwa, Rani:2016gnl, Rodrigues:2016uor, Mai:2017riq, Singh:2019ykp, Nashed:2020kjh, Bhatti:2018fsc, Ashraf:2020yyo, Ditta:2021wfl, Bahamonde:2021srr, Bahamonde:2022lvh}.

On the other hand, with the development of observational capabilities, scientists  can study the properties of gravity in the strong-field  regime, within increasing accuracy. For example, the Event Horizon Telescope (EHT) provided  the first photograph of a black hole \cite{EventHorizonTelescope:2019dse,  EventHorizonTelescope:2019ggy, EventHorizonTelescope:2020qrl}, and the LIGO-Virgo collaboration has detected the first gravitational wave signal \cite{LIGOScientific:2016aoc, LIGOScientific:2016lio}. Since these observations may reveal the black-hole features close to the event horizon, one may use them as a tool to probe modified gravities \cite{Cai:2018rzd, Yan:2019hxx, Li:2021mzq}.

In particular, binary black hole mergers are an important source of gravitational waves (GW), and the GWs emitted during the ringdown stage can be decomposed as the superposition of a series of quasinormal modes (QNM) of a perturbed Kerr black hole \cite{Nollert:1999ji, Konoplya:2011qq, Isi:2021iql}. QNMs are characteristic modes of the perturbation equations of a black hole solution \cite{Zerilli:1970se, Chandrasekhar:1975zza, Kokkotas:1999bd, Berti:2009kk, Hatsuda:2021gtn}, and can therefore reveal useful information about the corresponding spacetime geometry. This feature has promoted the search and analysis of QNM to be an important subject in the gravitational-wave literature. So far, many efforts have been made to develop numerical and semi-analytical methods to calculate the frequencies of QNM, such as the Wentzel-Kramers-Brillouin (WKB) approximation, numerical integration, continued fractions, Frobenius method and so on. In particular, it has been extensively studied in \cite{Konoplya:2003ii, Matyjasek:2017psv, Konoplya:2019hlu} that the calculation accuracy of QNMs sensitively depends on the order of the WKB approximation method. With the developments of the aforementioned theoretical approaches, one can use gravitational-wave observations to test General Relativity, examining for instance the validity of the ``no-hair theorem'' \cite{Dreyer:2003bv, Berti:2005ys, Isi:2019aib}, and use the predictions of QNM properties to constrain and classify modified gravity theories \cite{Cano:2021myl, Wang:2004bv, Blazquez-Salcedo:2016enn, Franciolini:2018uyq,Becar:2019hwk,Aragon:2020xtm, Liu:2020qia, Karakasis:2021tqx, Gonzalez:2022upu}. Furthermore, the QNM frequency can also partly reveal the stability of spacetime geometry under small perturbations \cite{Ishibashi:2003ap, Chowdhury:2022zqg}. Hence, the study of QNM has become a concerned topic \cite{Cai:2015fia, Cardoso:2019mqo, McManus:2019ulj, Guo:2021enm}.

In this work, we are interested in investigating the quasinormal modes in the case of black-hole solutions in $f(T)$ gravity. In particular, we first  extract perturbatively the vacuum spherically symmetric solutions, and then calculate the corresponding quasinormal modes frequency of a massless test scalar field, applying both the characteristic integration method 
\cite{Gundlach:1993tp} as well as the Wentzel-Kramers-Brillouin (WKB) approximation method \cite{Iyer:1986np}. The outline of the work is as follows. In Section \ref{BHfT},  we briefly review $f(T)$ gravity and extract the static spherically symmetric solutions applying the perturbative method. In Section \ref{sec: QNM calculation}, we obtain the QNM frequencies and the time evolution of the dominant mode. Finally, Section \ref{sec: conclusion} is devoted to discussion and conclusions.

\section{Black holes in $f(T)$ gravity}
\label{BHfT}

In this section, we first present the foundations and  field equations in $f(T)$ gravity and then solve them perturbatively, extracting the black hole solutions.

\subsection{$f(T)$ gravity}
\label{sec: theory review}

In teleparallel gravity, one uses the tetrad field as a dynamical variable which forms a set of orthonormal vectors in the tangent space, related to the metric tensor through 
\begin{equation}
    g_{\mu \nu}=\eta_{a b} h{^{a}}{_{\mu}} h{^{b}}{_{\nu}},
\end{equation}
where $h{^{a}}{_{\mu}}$ are the tetrad components and $\eta_{ab}=\text{diag}(+1,-1,-1,-1)$ the Minkowski metric. Throughout this article, Greek indices indicate spacetime coordinates, and Latin indices are used for tangent-space coordinates. 

While in  General Relativity one uses  the torsionless Levi-Civita connection, in teleparallel gravity one introduces the Weitzenb$\ddot{\text{o}}$ck connection, defined as:
\begin{equation}
    \Gamma{^\lambda}{_{\mu\nu}} \equiv h{_{a}}{^{\lambda}} (\partial_{\nu} h{^{a}}{_\mu}+\omega{^a}{_{b\mu}} h{^b}{_\nu}),
\end{equation}
with $\omega{^a}{_{b\mu}}$  the spin connection encoding the inertial spacetime effects, which is supposed to be zero when we transform to an inertial frame with gravity turned off. Thus, the spin connection can be calculated  as \cite{Krssak:2015oua,Krssak:2018ywd}:
\begin{equation}
    \omega{^{a}}{_{b \mu}}=\frac{1}{2} h_{(\mathrm{r}) \mu}^{c}\left[f{_{b}}{^{a}}{_{c}}\left(h_{(\mathrm{r})}\right)+f{_{c}}{^{a}}{_{b}}\left(h_{(\mathrm{r})}\right)-f{^{a}}{_{b c}}\left(h_{(\mathrm{r})}\right)\right],
    \label{spinconn}
\end{equation}
where the tetrad with subscript $(r)$ is the reference tetrad with gravity turned off smoothly, and $f{^a}{_{bc}}$ is related to the tetrad through 
\begin{equation}
    f{^{c}}{_{a b}}=h{_{a}}{^{\mu}} h{_{b}}{^{\nu}}\left(\partial_{\nu} h{^{c}}{_{\mu}}-\partial_{\mu} h{^{c}}{_{\nu}}\right).
\end{equation}
The corresponding torsion tensor  is defined as
\begin{equation}
    T{^{\lambda}}{_{\mu \nu}}=h{_{a}}{^{\lambda}}\left(\partial_{\mu} h{^{a}}{_{\nu}}-\partial_{\nu}  h{^{a}}{_{\mu}}+\omega{^{a}}{_{b \mu}} h{^{b}}{_{\nu}}-\omega{^{a}}{_{b \nu}} h{^{b}}{_{\mu}}\right).
\end{equation}
Hence, by introducing the contortion and superpotential tensors as
\begin{equation}
    K{^{\rho}}{_{\mu \nu}} \equiv \frac{1}{2}\left(T{_{\mu}}{^{\rho}}{_{\nu}}+T{_{\nu}}{^{\rho}}{_{\mu}}-T{^{\rho}}{_{\mu \nu}}\right),
\end{equation}
and
\begin{equation}
    S{_{\rho}}{^{\mu \nu}} \equiv \frac{1}{2}\left(K{^{\mu \nu}}{_{\rho}}+\delta_{\rho}^{\mu} T{^{\alpha \nu}}{_{\alpha}}-\delta_{\rho}^{\nu} T{^{\alpha \mu}}{_{\alpha}}\right),
\end{equation}
respectively, we can define the torsion scalar
\begin{equation}
    T=S{_{\rho}}{^{\mu \nu}} T{^{\rho}}{_{\mu \nu}}.
\end{equation}
By using $T$ as the Lagrangian, we can achieve the teleparallel equivalent of General Relativity (TEGR) since the two theories give rise to the same field equations (since $T$ and the standard Ricci scalar $R$ corresponding to the Levi-Civita connection  differs only by a boundary term).

One can extend TEGR to $f(T)$ gravity by upgrading the Lagrangian to a  non-linear function of $T$, namely
\begin{equation}
    S=\int d^{4} x \frac{h}{16 \pi G}f(T),
\end{equation}
with $h=\det\left(h^{a}{}_{\mu}\right)=\sqrt{-g}$,  which is a novel gravitational modification, different from $f(R)$ gravity (since a function of a boundary term in general is not a boundary term). Finally, varying the action with respect to the tetrad, the vacuum field equations for $f(T)$ gravity are found to be
\begin{equation}
    \label{eq: field equation}
    h\! \left(\! f_{TT} S{_{a}}{^{\mu \nu}} \partial_{\nu} T -f_{T} T{^{b}}{_{\nu a}} S{_{b}}{^{\nu \mu}}+f_{T} \omega{^{b}}{_{a \nu}} S{_{b}}{^{\nu \mu}}+\frac{1}{2} f h{_{a}}{^{\mu}}\! \right)+f_{T} \partial_{\nu}\left(h S{_{a}}{^{\mu \nu}}\right)=0,
\end{equation}
with $f_{T}\equiv\partial f/\partial T$, and
$f_{TT}\equiv\partial^{2} 
f/\partial T^{2}$.

\subsection{Spherically symmetric solutions}
\label{sec: metric solution}

Let us now proceed to the extraction of spherically symmetric solutions in $f(T)$ gravity, which is one of the main steps one can follow in order to extract information on the features of the theory  \cite{Will:1993ns}. Since it is known that a realistic $f(T)$ gravity should be a small deviation from TEGR \cite{Wu:2010mn,Cardone:2012xq,Nesseris:2013jea, Nunes:2016plz,Basilakos:2018arq, Xu:2018npu,Ren:2022aeo},  any $f(T)$ form can be efficiently approximated as 
\begin{equation}
    f(T)=T+\alpha T^2+{\cal{O}}(T^3),
\end{equation}
where the constant $\alpha$ describes the extent of modification. This is the simplest and most direct nonlinear extension of TEGR gravity and moreover it is the leading order contribution when considering the weak-field scenario. The above modification of TEGR, i.e., of GR, gives rise to black-hole solutions that perturbatively deviate from the Schwarzschild solution 
\cite{Bahamonde:2020bbc, DeBenedictis:2016aze, Bahamonde:2019zea, Ren:2021uqb}. These allow us to examine and constrain the theory with Solar System data \cite{Iorio:2012cm, Iorio:2015rla, Farrugia:2016xcw} or through galaxy lensing observations \cite{Chen:2019ftv}. In the following we will apply this approximation to demonstrate the possibility of testing and constraining $f(T)$ gravity with QNM observations.

We  consider the general metric  
\begin{equation}
    ds^2 =A(r)^2dt^2-B(r)^2dr^2-r^2d\Omega^2,
\end{equation}
and thus the corresponding tetrad can be written as
\begin{equation}
    \label{eq: tetrad ansatz}
    h^{a}{}_{\mu}=
    \left(\begin{array}{cccc}
        A(r) & 0 & 0 & 0 
        \\ 
        0 & B(r) \sin (\theta) \cos (\phi) & r \cos (\theta) \cos (\phi) & -r \sin (\theta) \sin (\phi) 
        \\
        0 & B(r) \sin (\theta) \sin (\phi) & r \cos (\theta) \sin (\phi) & r \sin (\theta) \cos (\phi) 
        \\
        0 & B(r) \cos (\theta) & -r \sin (\theta) & 0
    \end{array}\right).
\end{equation}
Using (\ref{spinconn}) one can easily find that the spin connection associated with this tetrad is zero \cite{Krssak:2018ywd}, which is consistent with the fact that this tetrad transforms to an inertial frame when gravity is turned off \cite{Pfeifer:2021njm}. In this case, the general field equations \eqref{eq: field equation} become
\begin{eqnarray}
&&
    \!\!\!\!\!\!\!\!\! 
    A(r) \left[4 r f_T B'(r)  +f r^2 
    B(r)^3+4 B(r)^2 \left(r f_{\text{TT}} T'+f_T\right)\right. \notag
    \\ 
&& 
    \left. 
    -4 B(r) \left(r 
    f_{\text{TT}} T'+f_T\right)\right]
    +4 r (B(r)-1) B(r) f_T A'(r) =0,
    \label{eq11}
    \\
&&
    \!\!\!\!\!\!\!\!\! 
    A(r) \left[f r^2 B(r)^2+4 B(r) f_T-4 
    f_T\right]
    + 4 r [B(r)-2] f_T A'(r) =0,
    \\
&&
    \!\!\!\!\!\!\!\!\! 
    A(r) \left[B(r)^3 \left(f r^2-2 f_T\right)+2 B(r)^2 \left(r f_{\text{TT}} T'+2 f_T\right)-2 B(r) \left(r f_{\text{TT}} T'+f_T\right)\right.\notag
    \\
&&
    \left.
    +2 r f_T B'(r)\right]
   -2 r \left\{B(r) \left[r f_T A''(r)+A'(r) \left(r f_{\text{TT}} T'+3 f_T\right)\right]\right.\notag
   \\
&&
    \left.
    -r f_T A'(r) B'(r)-2 B(r)^2 f_T A'(r)
    \right\} 
    =0.
    \label{eq33}
\end{eqnarray}
Assuming that the solution is close to the Schwarzschild, one we can solve these field equations with the perturbative method, and in order to obtain the solution  we need to determine the expression of the metric perturbation. In the following discussion we will consider  two different ans$\ddot{\text{a}}$tze, called ``ansatz1'' and ``ansatz2'' \cite{Mann:2021mnc}. Finally, for convenience we will also re-express the metric as
\begin{equation}
    \label{eq: N and F}
    ds^2=N(r)dt^2-\frac{dr^2}{F(r)}-r^2d\Omega^2,
\end{equation}
where $N(r)=A(r)^2$ and $ F(r)=1/B(r)^2$.

\begin{figure}[ht]
    \centering
    \includegraphics[width=0.45\textwidth]{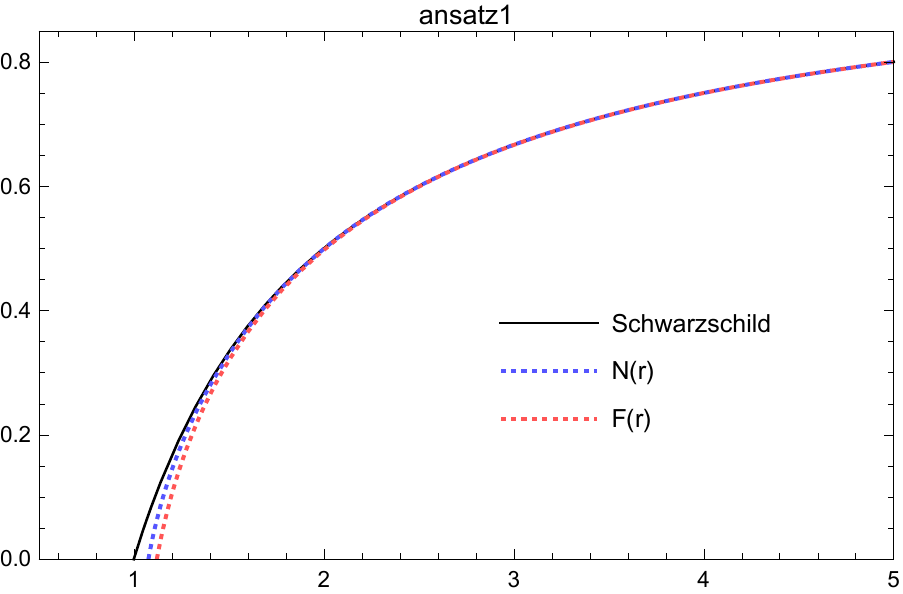}
    \includegraphics[width=0.45\textwidth]{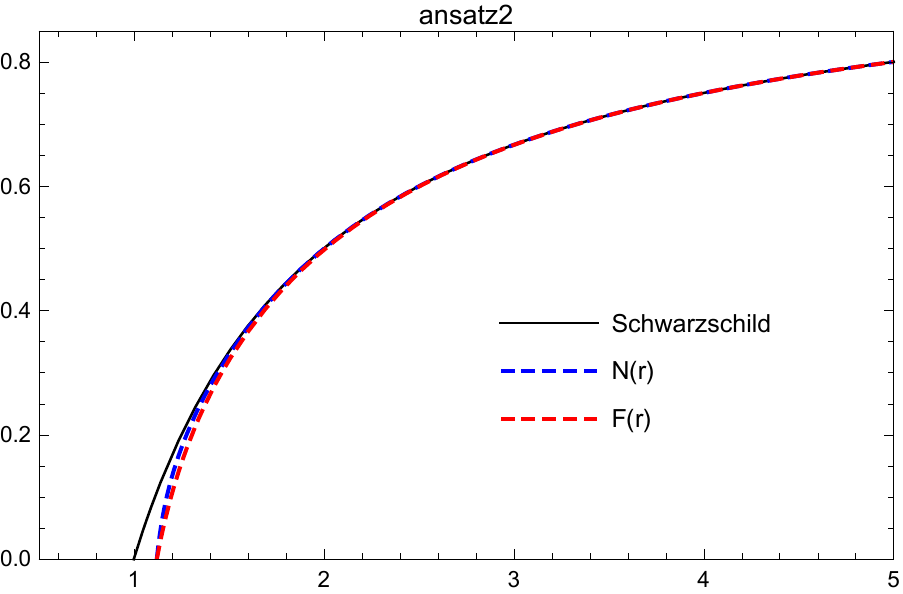}
    \caption{\it{The perturbative solution for the metric in $f(T)$ gravity black holes, for ansatz1 of \eqref{Ansatz1ab} with $\alpha=0.05$ (upper graph) and ansatz2  of \eqref{Ansatz2ab} with $\alpha=0.05$  (lower graph), in units where  $2M=1$.}}
    \label{fig:perturbation}
\end{figure}

\subsubsection{Ansatz 1}

The perturbed metric can be written as a small deviation from Schwarzschild metric, parameterized with the functions \cite{Bahamonde:2020bbc, DeBenedictis:2016aze, Bahamonde:2019zea, Pfeifer:2021njm}
\begin{align}
&
    A(r)=\sqrt{1-\frac{2 M}{r}+\varepsilon  a(r)},\nonumber
    \\
&
    B(r)=\sqrt{\frac{1}{1-\frac{2 M}{r}+\varepsilon  b(r)}},
    \label{Ansatz1ab}
\end{align}
where $a(r)$ and $b(r)$ are small functions compared to $1-\frac{2M}{r}$, and $\varepsilon$ is the tracking parameter of the perturbation. Inserting it into the field equations \eqref{eq11}- \eqref{eq33} and keeping terms up to first order in  $\varepsilon$, we obtain the system of equations
\begin{eqnarray}
&& 
    \!\!\!\!\!\!\!\!\!\!\!\!\! 
    r^3 b'(r)+r^2 b(r)-2 \alpha x^{-3} (x-1)^3 \left(x^2+5 x+10\right) =0,
    \\
&& 
    \!\!\!\!\!\!\!\!\!\!\!\!\! 
    r^3 x^2 a'(r)-r^2 \left(x^2-1\right) x^2 a(r)+2 \alpha +r^2 x^4 b(r)+2 \alpha  x^4-8 \alpha  x^3+12 \alpha  x^2-8 \alpha  x =0,
    \\
&& 
    \!\!\!\!\!\!\!\!\!\!\!\!\!   
    2 r^4 x^3 a''(r)-r^3 x^5 a'(r)+3 r^3 x^3 a'(r)+r^2 \left(x^4-1\right) x^3 a(r)+20 \alpha +r^3 x^3 b'(r)
   \notag\\
&&
    \!\!\!\!\!\!\!\!\! 
   -r^2 \left(x^4-1\right)x^3 b(r)+4 \alpha x^6-20 \alpha  x^4+60 \alpha  x^2-64 \alpha  x =0,
\end{eqnarray}
where $x\equiv\sqrt{\frac{r}{r-2M}}$. Thus, we can extract the analytic solutions for the functions $a(r)$ and $b(r)$  as
\begin{eqnarray}
&& 
    \!\!\!\!\!\!\!\!\! \!\!\!\!\!\!\!\!\!  
    a(r)=c_2\left(1\!-\!\frac{2 M}{r}\right)+\frac{c_1}{r}-\frac{\alpha }{3}r^{-2} x^{-2} \left(x^2\!-\!1\right)^{-2}
    \nonumber\\
&&
    \, \cdot    \Big[-51 x^6+93 x^4+128 x^3-45 x^2 -12 \left(x^2\!-\!3\right) x^4 \log (x)+3\Big],
    \\
&& 
    \!\!\!\!\!\!\!\!\! \!\!\!\!\!\!\!\!\!   
    b(r)=\frac{c_1}{r}-\frac{\alpha}{3} r^{-2} x^{-3} \left(x^2\!-\!1\right)^{-1} 
    \nonumber\\
&&
    \, \cdot \Big[63 x^5+12 x^5 \log (x)-24 x^4 +12 x^3+64 x^2-75 x+24\Big], 
\end{eqnarray}
where the constants $c_1$ and $c_2$ are chosen to be $c_1=\frac{32 \alpha }{3 M}$ and $c_2=-\frac{32 \alpha }{3 M^2}$ in order to be consistent with the asymptotic behavior of Schwarzschild solution. Hence, for ansatz1 we finally obtain  the perturbative result 
\begin{align}
    N(r)=
&
    1-\frac{2 M}{r}-\frac{\alpha  \left[13 x^6-99 x^4+128 x^3-45 x^2-12 \left(x^2-3\right) x^4 \log (x)+3\right]}{3 r^2 x^2 \left(x^2-1\right)^2},
    \\
    F(r)=
&
    1-\frac{2 M}{r}+\frac{\alpha  \left[x^5-12 x^5 \log (x)+24 x^4-12 x^3-64 x^2+75 x-24\right]}{3 r^2 x^3 \left(x^2-1\right)},
\end{align}
which quantifies the deviation from the Schwarzschild black hole. This result is in consistent with previous results \cite{DeBenedictis:2016aze, Bahamonde:2019zea, Ren:2021uqb}, etc. In order to present the results in a more transparent way, in the upper graph of  Fig. \ref{fig:perturbation} we depict the metric functions for the case of ansatz1, while for comparison we add the Schwarzschild case, too (note that in $f(T)$ case $ F(r)$ and $N(r)$ are not the same).

\subsubsection{Ansatz 2}

The perturbed metric can be alternatively  parameterized as \cite{Mann:2021mnc}
\begin{align}
&
    \text{A}(r)=\sqrt{e^{2 \varepsilon  a(r)} \left[1-\frac{2 M+\varepsilon b(r)}{r}\right]},
    \nonumber\\
&
    \text{B}(r)=\frac{1}{\sqrt{1-\frac{2 M+\varepsilon  b(r)}{r}}}.
    \label{Ansatz2ab}
\end{align}
Note that within  this ansatz  the two metric functions $N(r)$ and $F(r)$ have the same root. Inserting into \eqref{eq11}- \eqref{eq33} we finally extract the analytic solutions for $a(r)$ and $b(r)$ as  
\begin{eqnarray}
&&
    \! \!\!\!\!\!\!\!\!\!\!\! \!\!\!\!
    a(r)=c_1+ \frac{\alpha}{3}  r^{-2} x^{-1}\left(x^2\!-\!1\right)^{-2}   
    \nonumber\\
&& 
    \cdot\Big[6 x^7-12 x^6+21 x^5+12 x^5 \log (x)+108 x^4-66 x^3-20 x^2+39 x-12\Big],
    \\
&& 
    \!\!\!\!\!\!\!\!\!\!\!\! \!\!\!\! 
    b(r)=c_2-\frac{\alpha}{3}   r^{-1} x^{-3} \left(x^2\!-\!1\right)^{-1} 
    \nonumber\\
&&
    \cdot\Big[12 x^5 \log (x) +63 x^5  -24 x^4 +12 x^3+64 x^2-75 x+24 \Big].
\end{eqnarray}
with  $c_1= -\frac{16 \alpha }{3 M^2}$ and $c_2= -\frac{32 \alpha }{3 M}$ in order to acquire the asymptotic behavior of the Schwarzschild solution. Hence, under ansatz2 the metric functions are found to be    
\begin{align}
    N(r)=
&
    \exp \left\{-\frac{2\alpha  \left[-6 x^6+12 x^5+43 x^4-12 x^4 \log (x)-108 x^3+66 x^2+20 x+\frac{12}{x}-39\right]}{3 r^2 \left(x^2-1\right)^2}\right \}
    \notag\\
&
    \cdot\frac{\left\{(x-1) \left[3 r^2 x (x+1)+\alpha  \left(x^4+25 x^3+13 x^2-51 x+24\right)\right]-12 \alpha  x^5 \log (x)\right\}}{3 r^2 x^3\left(x^2-1\right)},
   \\
   F(r)=
&
    \frac{(x-1) \left[3 r^2 x (x+1)+\alpha  \left(x^4+25 x^3+13 x^2-51 x+24\right)\right ]-12 \alpha  x^5 \log (x)}{3 r^2 x^3 \left(x^2-1\right)}.
\end{align}
In the lower graph of Fig. \ref{fig:perturbation} we show the metric functions for the case of ansatz2, on top of the Schwarzschild result. 
 
In conclusion, we see that the two metric ans$\ddot{\text{a}}$tzes give slightly different metric solutions, which are both close to the Schwarzschild case when we are outside the horizon. Nevertheless, as one will see in the next section, these small differences (and the fact that for ansatz2 the functions $N(r)$ and $F(r)$ have the same root) lead to differences in the QNM frequency of the system.

\section{Quasinormal modes analysis}
\label{sec: QNM calculation}

In this section we proceed to the main analysis of the present work, namely to the calculation of the Quasinormal modes (QNM) frequencies of the above black-hole solutions, applying  both the characteristic integration method and the WKB approximation method \cite{Cai:2015fia}.

\subsection{Schr$\ddot{\text{o}}$dinger-like equation and effective potential}

The motion of a scalar particle $\psi$ is determined by the Klein-Gordon equation: 
\begin{equation}
    \square \psi=0,
\end{equation}
where the covariant derivative is defined in terms of the  Levi-Civita connection. In the case of a spherically symmetric background,  with the metric functions given in \eqref{eq: N and F} and the corresponding explicit calculation of the box operator, the Klein-Gordon equation becomes
\begin{eqnarray}
&&  
    \!\!\!\!\!\!\!\!
    \left(\frac{2 F}{r}+\frac{\partial_{r} F}{2}+\frac{F \partial_{r} N}{2 N}\right) \partial_{r} \psi+F \partial_{r}^{2} \psi -\frac{\partial_{t}^{2} \psi}{N}+\frac{\partial_{\phi}^{2} \psi}{r^{2} \sin ^{2} \theta}+\frac{\partial_{\theta} \psi}{r^{2} \tan \theta}+\frac{\partial_{\theta}^{2} \psi}{r^{2}}=0.
\end{eqnarray}
As usual, we proceed by considering the scalar field to obey the separated variable form
\begin{equation}
    \psi(t, r, \theta, \phi)=\sum_{l m} \frac{1}{r} \Psi_{l}(r) Y_{l m}(\theta, \phi) e^{-i \omega t},
\end{equation}
where $Y_{lm}(\theta,\phi)$ are the spherical harmonic functions. Therefore, the radial part of the Klein-Gordon equation becomes
\begin{align}
&&  
    \!\!\!\!\!\!\!\!\left[\frac{r^{2} \omega^{2}}{N}-\frac{r \partial_{r} F}{2}-\frac{r F \partial_{r} N}{2 N}-l(l+1)\right] \Psi_{l}+\left(\frac{r^{2} \partial_{r} F}{2}+\frac{r^{2} F \partial_{r} N}{2 N}\right) \partial_{r} \Psi_{l}+r^{2} F \partial_{r}^{2} \Psi_{l}=0.
\end{align}
In order to simplify this equation, we can transform the radial coordinate $r$ to the ``tortoise coordinate'' $r^*$ through the variable transformation
\begin{equation}
    d r^{*}=d r / \sqrt{F(r) N(r)},
\end{equation}
where $r^{*}$ is similar to the tortoise coordinate defined in Schwarzschild metric \cite{Cai:2015fia}. We mention that this tortoise coordinate is defined only outside the event horizon, 
namely the root of the metric function $F(r)$. In this way, the radial part of the scalar-field equation of motion can be written as a Schr$\ddot{\text{o}}$dinger-like equation
\begin{equation}
    \label{eq: Schrodinger-like equation}
    \left[\partial_{r *}^{2}+\omega^{2}-V_{l}(r^*)\right]\Psi_{l}\left(r^{*}\right)=0,
\end{equation}
with the effective potential $V_l(r)$ in the form of
\begin{equation}
    \label{eq: effective potential}
    V_{l}(r)=V_{l}\left(r\left(r^*\right)\right)=\frac{F \partial_{r} N+N \partial_{r} F}{2 r}+\frac{l(l+1) N}{r^{2}}.
\end{equation}

Similarly, one can get the Schr$\ddot{\text{o}}$dinger-like equation and effective potential for an electromagnetic field. In this case, the electromagnetic potential $A_\mu$ can be expressed in terms of four-dimensional vector spherical harmonics as follows,
\begin{align}
	A_\mu (t,r,\theta,\phi)
	=\sum_{lm}
	\left[
		\left(
		\begin{gathered}
			0 \\
			0 \\
			\frac{a^{lm}(t,r)}{\sin\theta}\partial_\phi Y^{lm} \\
			-a^{lm}(t,r) \sin\theta \partial_\theta Y^{lm}
		\end{gathered}
		\right)+
		\left(
		\begin{gathered}
			f^{lm}(t,r) Y^{lm} \\
			h^{lm}(t,r) Y^{lm} \\
			k^{lm}(t,r) \partial_\theta Y^{lm} \\
			k^{lm}(t,r) \partial_\phi Y^{lm}
		\end{gathered}
		\right)
	\right].
\end{align}
Note that, the first term in the right-hand side has parity $(-1)^{l+1}$ and the second term has parity $(-1)^{l}$. Then, we plug these expressions into Maxwell equations $\nabla_\nu F^{\mu\nu}=0$ where $F_{\mu\nu}=\nabla_\mu A_\nu-\nabla_\nu A_\mu$.
%\begin{equation}
%    \nabla_\nu F^{\mu\nu}=0 ~, \quad \text{where} \quad
%    F_{\mu\nu}=\nabla_\mu A_\nu-\nabla_\nu A_\mu.
%\end{equation}
After performing proper coordinate transformation, the integration condition of the Maxwell equations can yield
\begin{equation}
    \label{eq: master equation_alm}
    \frac{\partial^2 a^{lm}}{\partial r^{*2}}-\partial_0^2 a^{lm}-N\frac{l(l+1)}{r^2} a{^{lm}} ~,
\end{equation} 
for parity $(-1)^{l+1}$ section, and
\begin{equation}
    \label{eq: master equation_blm}
    \frac{\partial^2 b^{lm}}{\partial r^{*2}}-\partial_0^2 b^{lm}-N\frac{l(l+1)}{r^2} b{^{lm}} ~,
    \quad \text{where} \quad
    h{^{lm}}_{,0}-f{^{lm}}{_{,r}}=\frac{\ell(\ell+1)}{r^2}\sqrt{\frac{N}{F}}b{^{lm}},
\end{equation}
for parity $(-1)^{l}$ section. These equations \eqref{eq: master equation_alm} and \eqref{eq: master equation_blm} can be summarized in the 
Schr$\ddot{\text{o}}$dinger-like equation \eqref{eq: Schrodinger-like equation} with effective potential
\begin{equation}
    V(r)=N(r)\frac{l(l+1)}{r^2} ~.
\end{equation}

Analogous treatment for fields of spin $s$, leads to similar Schr$\ddot{\text{o}}$dinger-like equation with effective potential of form \cite{Cardoso:2001bb, Konoplya:2022tvv}
\begin{equation}
    V(r)=N(r)\frac{l(l+1)}{r^2}+\frac{1-s}{2r}\frac{d}{dr}(N(r)F(r)) ~.
\end{equation}
To be specific, we focus our analyses on the QNMs of a massless scalar field for demonstration, and also give the QNM frequencies results for a massless vector field, i.e., the electromagnetic field, while leave the fields of other spins for a future consideration. The gravitational perturbations would be rather complicated and may be addressed in certain concrete cases.

\begin{figure}
    \centering
    \includegraphics[width=0.55\textwidth]{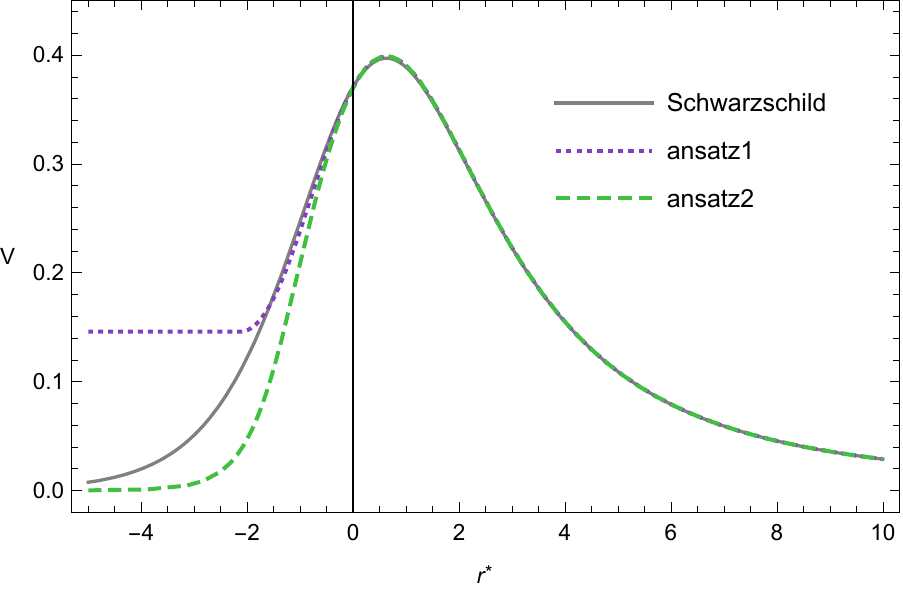}
    \caption{\it{The effective potential \eqref{eq: effective potential} as a function of the tortoise coordinate $r^*$, corresponding to the metric solutions of subsection \ref{sec: metric solution}, arising from the two imposed ans$\ddot{\text{a}}$tze, for orbital angular momentum  $l=1$, and $\alpha=0.05$ in units where $2M=1$}. The dashed purple line corresponds to ansatz1, the dashed green line corresponds to ansatz2, and for completeness we add the Schwarzschild result too (solid grey line).}
    \label{fig:potential}
\end{figure}

The effective potential plays an important role in the QNM calculation. In order to provide a more transparent picture, in Fig. \ref{fig:potential} we depict its form of a massless scalar field for the two extracted solutions of the previous section, corresponding to the two imposed ans$\ddot{\text{a}}$tzes. As we can observe, there is a distinct difference in the shape of the effective potentials in the two cases, although the two metric solutions are nearly indistinguishable for radial distances larger than 1.5 Schwarzschild radius (see Fig. \ref{fig:perturbation}). The major difference appears at negative $r^{*}$, where ansatz2 exhibits a larger slope, while ansatz1 acquires a non-zero asymptotic value. This is due to the fact that the metric functions $N(r)$ and $F(r)$ for ansatz1 have different roots, and thus the effective potential obtains a non-zero value at the event horizon, which is not the case for  ansatz2. This difference will affect the evolution of the scalar field, as we will see below.

\subsection{Characteristic integration method}

\begin{figure} [ht]
    \centering
    \includegraphics[width=0.55\textwidth]{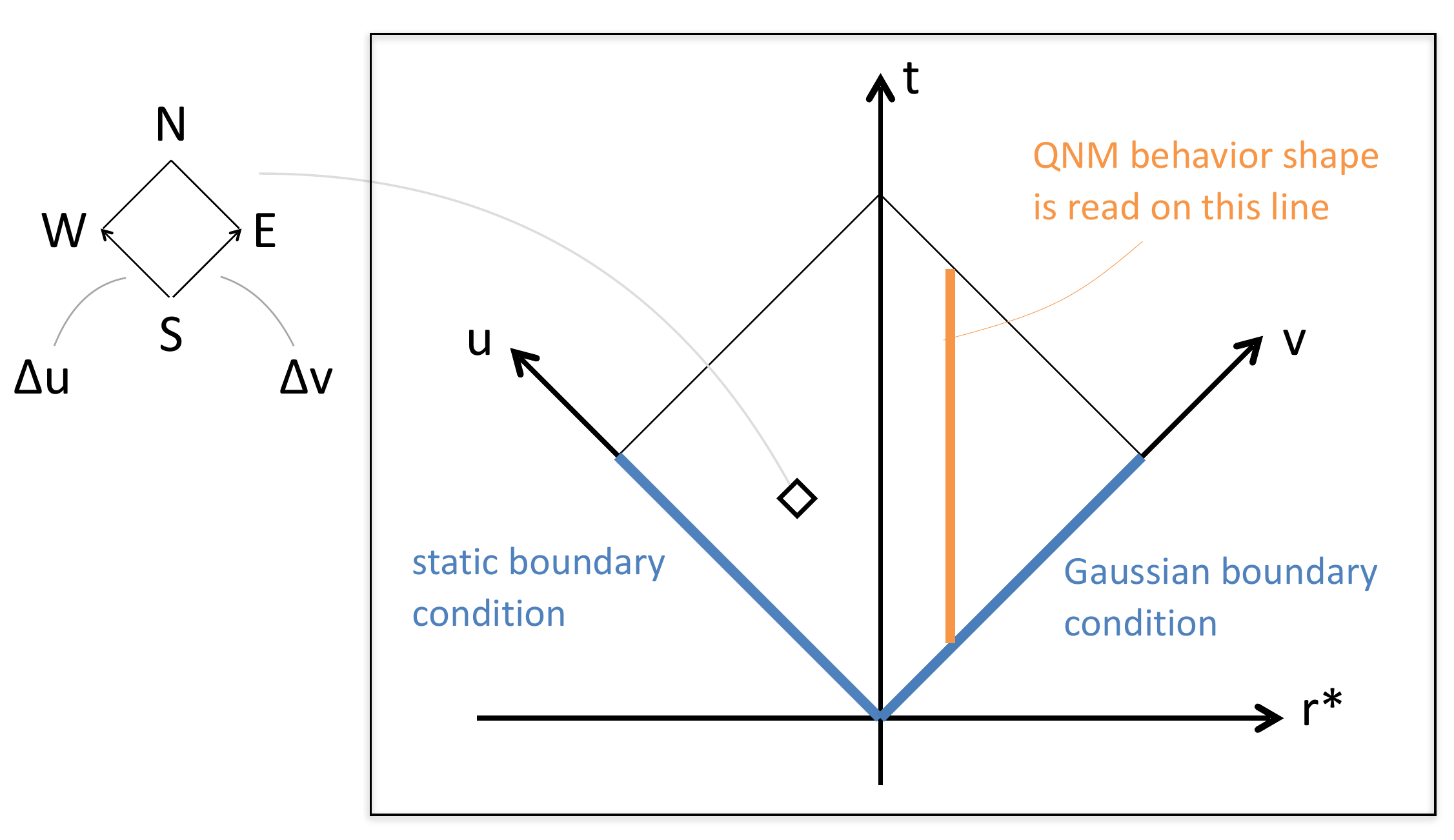}
    \caption{\it{Demonstration graph for the characteristic integration method.}}
    \label{fig:discretize}
\end{figure}

We first calculate the QNM frequency applying the characteristic integration method \cite{Gundlach:1993tp}. In this method, we discretize the equation of motion, and then solve numerically for the evolution of the scalar field, under specific initial and boundary conditions, in order to extract the dominant QNM frequency from the form of the above time evolution. To implement the numerical calculation, we introduce the light-cone coordinates
\begin{equation}
    u=t-r^{*}, \quad v=t+r^{*}.
\end{equation}
In these coordinates, the Schr$\ddot{\text{o}}$dinger-like equation \eqref{eq: Schrodinger-like equation} is written as
\begin{equation}
    -4 \frac{\partial^{2} \Phi_{l}(u, v)}{\partial v \partial 
u}-V_{l}(u(r),v(r)) \Phi_{l}(u, v)=0.
\end{equation}
We proceed by discretizing this equation as
\begin{eqnarray}
&& 
   \!\!\!\!\!\!\!\!\!\!\!\!\!\!\!\!\!\!\!\!\!\!\!
    \Phi_{l}(N)=\Phi_{l}(W)+\Phi_{l}(E)-\Phi_{l}(S) -\frac{q^{2}}{8} V(S)\left[\Phi_{l}(W)+\Phi_{l}(E)\right]+\mathcal{O}\left(q^{4}\right), 
\end{eqnarray}
with $S=(u, v)$, $W=(u+h, v)$, $E=(u, v+h)$, $N=(u+h, v+h)$,  and where we have introduced the variable $\Phi=\Psi e^{-i \omega t}$. In this calculation, $h$ is the step length, set to   $h=0.1$. The schematic diagram for the numerical implementation is presented in Fig. \ref{fig:discretize}. As shown in the graph, we require initial and boundary conditions on the null 
boundary $u=u_0$ and $v=v_0$ \cite{Gundlach:1993tp}. After integrating over the square, we can read the evolution of the scalar field at constant radius. Hence, through Fourier transformation or function fit \cite{McManus:2019ulj}, we can extract the corresponding QNM frequency of the system. 

\begin{figure}[ht]
    \centering
    \includegraphics[width=0.55\textwidth]{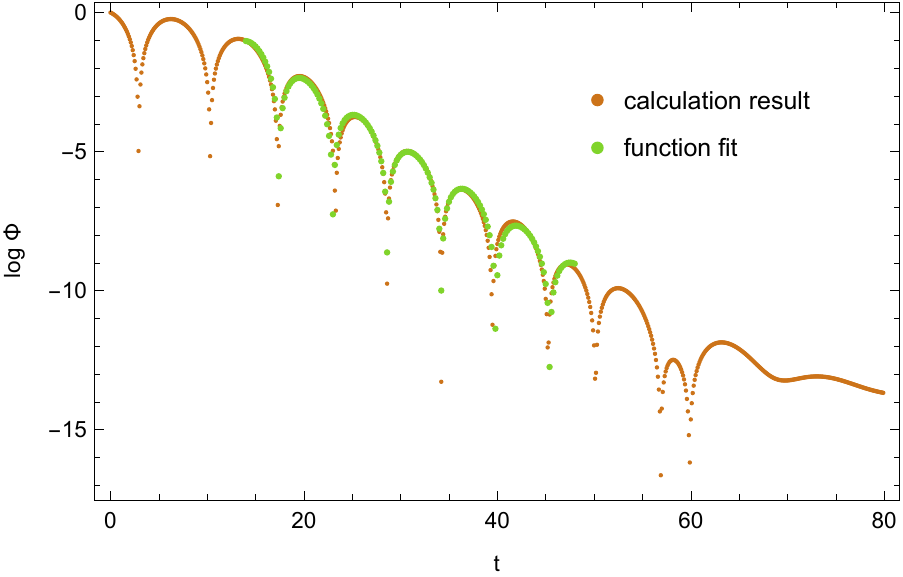}
    \caption{\it{Decay of the massless scalar field mode function $\log\left|\Phi(t)\right|$ for a black hole in $f(T)$ gravity,  for the case of ansatz2, for $l=1$ and for $\alpha=0.05$ in units of $2M=1$.}}
    \label{fig: shape example}
\end{figure}

In Fig. \ref{fig: shape example} we provide  an example of the numerical computation and function fit of the QNM form. We choose $\alpha=0.05$, in order for the deviation from Schwarzschild 
solution to be small, and we perform the calculation for massless scalar field and the case of ansatz2, choosing $l=1$ for demonstration. The orange curve corresponds to the numerical result of the evolution for a mode function on a logarithmic scale.  From this graph  we can see that the evolution of the scalar field is firstly dominated by the quasinormal vibrations, and then it decays with a power-law tail. This is similar to the case of the Schwarzschild solution, and it is the standard time domain profile of black holes. Additionally, the green line is the function fit of the numerical data, with function $\log \left|a \exp [-i(b+c i)(t-d)]\right|$. As we  can see, the function fit is efficient. Once we extract the function fit parameters $a$, $b$, $c$ and $d$, we can read the real and imaginary part of the QNM frequency from the parameters $b$ and $c$. In the following, we use this function fitting method in order to obtain the QNM frequencies.

\begin{figure}[ht]
    \centering
    \includegraphics[width=0.95\textwidth]{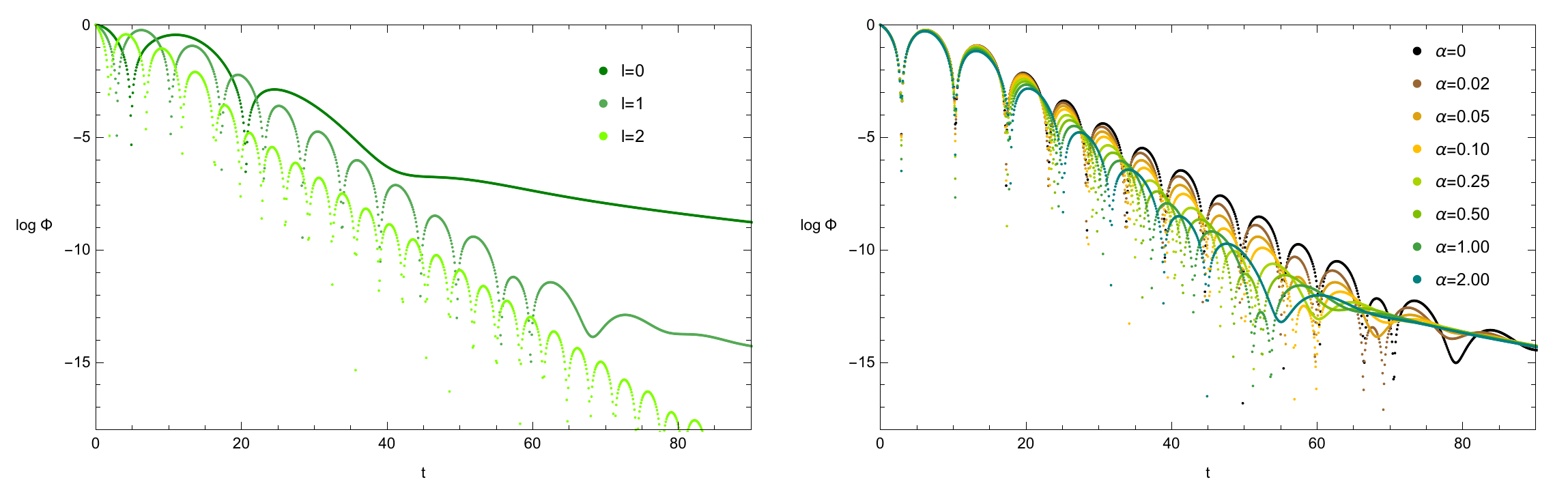}
    \caption{\it{ Decay of the massless scalar field mode function  $\log\left|\Phi(t)\right|$ for a black hole in $f(T)$ gravity,  for the case of ansatz2, for varying orbital angular momentum $l$ and fixed $\alpha=0.05$ in units of $2M=1$ (upper graph), and for varying $\alpha$ and fixed $l=1$ (lower graph).}}
    \label{fig: shape parameter}
\end{figure}
\begin{figure*}
    \centering
    \includegraphics[width=1.0\textwidth]{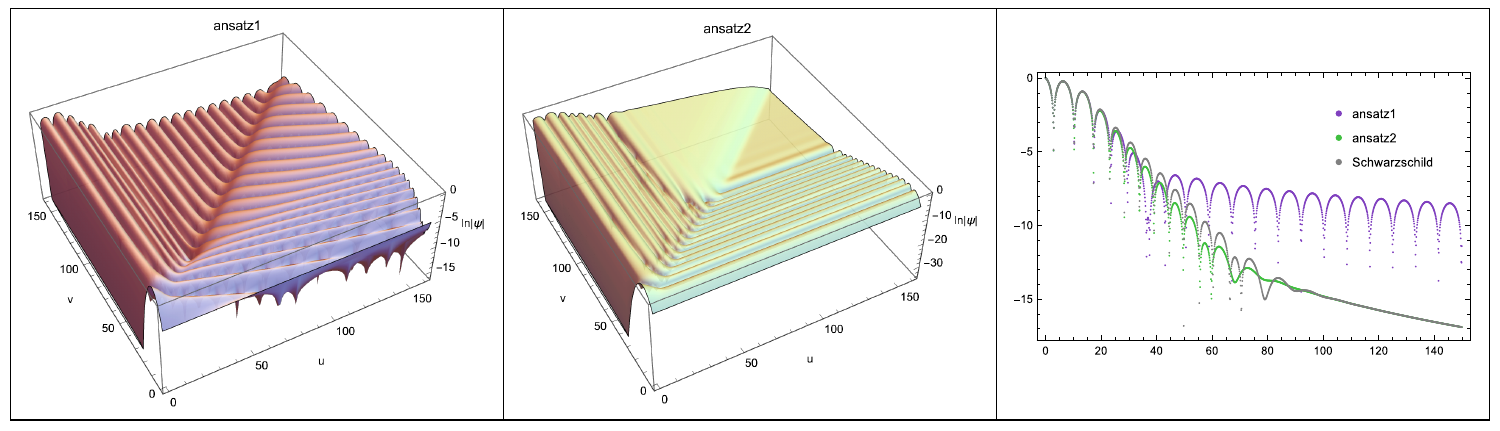}
    \caption{\it{Decay of the field mode function $\log\left|\Phi(t)\right|$ for a black hole in $f(T)$ gravity, for both ansatz1 and ansatz2, for $l=1$ and for $\alpha=0.05$ in units of $2M=1$. The left and the middle graphs depict the time evolution of the scalar field in the two ans$\ddot{\text{a}}$tze with respect to $u$ and $v$, while the right graph presents the time evolution for constant $r^{*}$.  }}
    \label{fig: shape ansatz}
\end{figure*}
We proceed by investigating the effect of the $f(T)$ model parameter $\alpha$, as well as of the orbital angular momentum $l$, on the field decay behavior. Focusing on metric ansatz2, in the upper graph of  Fig. \ref{fig: shape parameter} we present the field decay for different $l$, varying from 0 to 2, with a massless scalar field and $\alpha$ fixed to 0.05. As we can see, for larger $l$  we have smaller period of quasinormal vibration, a  behavior which is similar to the Schwarzschild case. Furthermore, in the lower graph of Fig. \ref{fig: shape parameter}, we depict the field decay for different $\alpha$, with fixed $l=1$. The $\alpha=0$ curve corresponds to the Schwarzschild case, while the $\alpha\neq0$ curves present the effect of the $f(T)$  modification. We can see that the field decay behavior changes smoothly for different $\alpha$. Moreover, we observe that for increasing  $\alpha$ we have longer period and larger slope. This may be comprehensible since under the $f(T)$ modification the metric solution has larger slope around the event horizon (see Fig.  \ref{fig:perturbation}), and thus larger slope for the field decay.

Finally, let us compare the cases of ansatz1 and ansatz2. In Fig. \ref{fig: shape ansatz} we depict the decay of the massless scalar field mode function in the two cases, for $\alpha=0.05$ and $l=1$. In order to present the comparison in a more transparent way, we display both the 3D evolution, as well as the time evolution at constant radius. As we observe, the field decay for ansatz2 is similar to the standard case of Schwarzschild metric, while for ansatz1 there is no typical power-law tail. This results from the fact that in ansatz1 the metric functions $F(r)$ and $N(r)$ have different roots, and thus the effective potential acquires a  non-zero value at the event horizon, making the perturbation of the scalar field not to smooth out   as in the Schwarzschild case. Lastly, note that there are also some differences in the form and slope of the various curves in the lower graph, and this leads to different results for the QNM frequencies.

\subsection{WKB approximation method}

In order to calculate the QNM frequency semi-analytically, we can use the WKB approximation method \cite{Iyer:1986np,Molina:2003dc}. 
If we define $Q=\omega^2-V$, then the Schr$\ddot{\text{o}}$dinger-like equation \eqref{eq: Schrodinger-like equation} acquires the form 
\begin{equation}
    \frac{d^{2} \Psi_{l}}{d r^{* 2}}+Q(r^*)\Psi_{l}=0.
\end{equation}
For a resonant normal mode of the black hole, whose response to an external perturbation is maximal, we have the boundary condition
\begin{equation}
    \lim _{r^{*} \rightarrow \pm \infty} \Psi_{l} e^{i\omega r^{*}}=1.
\end{equation}

This semi-analytic method has been widely used in QNM frequency estimation. Incorporating up to N order terms, the square of the frequencies can be expressed as
\begin{equation}
    \omega^2=V\left(x_0\right)-i\left(n+\frac{1}{2}\right) \sqrt{2 Q_0^{\prime \prime}} \varepsilon-i \sqrt{2 Q_0^{\prime \prime}} \sum_{i=2}^N \varepsilon^j \Lambda_j,
    \label{eq: WKB formula}
\end{equation}
where $n$ is the overtone number, $\epsilon$ is a tracking parameter, $x_0$ is the position of the peak of the effective potential, and all the terms are calculated at $x_0$. The first two correction terms are given by \cite{Iyer:1986np}
\begin{align}
    \Lambda_2(n)
&
    =\frac{1}{(2Q_0^{''})^{1/2}}\left[\frac{1}{8} \frac{V_{0}^{(4)}}{V_{0}^{(2)}}\left(\frac{1}{4}+\xi^{2} \right)-\frac{1}{288}\left[\frac{V_{0}^{(3)}}{V_{0}^{(2)}}\right]^{2}\left(7+60 \xi^{2}\right)\right]
    \\
    \Lambda_3(n)
&
    =\frac{n+\frac{1}{2}}{2Q_0^{''}}\left[\frac{5}{6912} \frac{V_{0}^{(3) 4}}{V_{0}^{(2) 4}}\left(77+188 \xi^{2}\right)
    -\frac{1}{384} \frac{V_{0}^{(3) 2} V_{0}^{(4)}}{V_{0}^{(2) 3}}\left(51+100 \xi^{2}\right)\right. \notag
    \\ 
&
    \left.+\frac{1}{2304} \frac{V_{0}^{(4) 2}}{V_{0}^{(2) 2}}\left(67+68 \xi^{2}\right)
    +\frac{1}{288} \frac{V_{0}^{(3)} V_{0}^{(5)}}{V_{0}^{(2) 2}}\left(19+28 \xi^{2}\right)
    -\frac{1}{288} \frac{V_{0}^{(6)}}{V_{0}^{(2)}}\left(5+4 \xi^{2}\right)\right],
\end{align}
where $\xi=n+\frac{1}{2}$. Higher order correction terms can be found in Refs. \cite{Konoplya:2003ii, Matyjasek:2017psv}.

However, the convergence of WKB series is only asymptotical, and when the order is not high enough, the results may get worse as the order increases \cite{Konoplya:2019hlu, Matyjasek:2019eeu}. To enhance the accuracy, the Taylor expansion in \eqref{eq: WKB formula} can be replaced by the Pade approximation \cite{Matyjasek:2017psv}
\begin{equation}
    P_{\tilde{m}}^{\tilde{n}}(\varepsilon)=\frac{\sum_{n=0}^{\tilde{n}} A_n \varepsilon^n}{\sum_{n=0}^{\tilde{m}} B_n \varepsilon^n},
\end{equation}
which is known to give satisfactory results sometimes even out of the range it's mathematically proven to converge. Existing literature about Pade approximation showed that it can greatly improve the WKB method to get better results \cite{Matyjasek:2017psv, Konoplya:2019hlu}, especially for extremely high order Pade approximation \cite{Hatsuda:2019eoj, Matyjasek:2019eeu}, which can successfully reproduce many known results of accurate numerical calculations to high decimal places \cite{Matyjasek:2021xfg}. In our discussion, we use $P_6^6$ approximation to improve our WKB method calculation.

\begin{figure}[ht]
    \centering
    \includegraphics[width=3.1in]{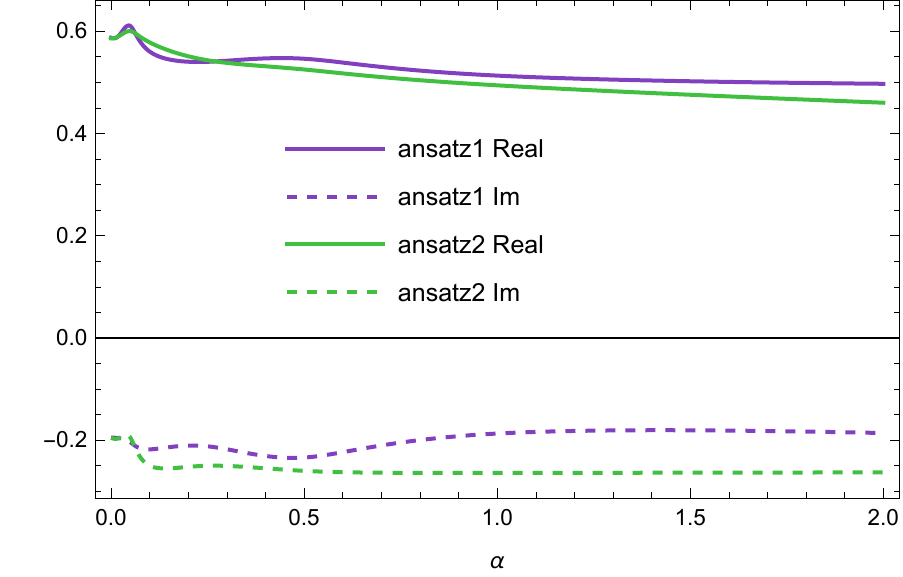}
    \caption{\it{The real and imaginary parts of the QNM frequency for a massless scalar field with respect to the model parameter $\alpha$, for $l=1$,  in the case of the WKB approximation method. }}
    \label{fig: WKB}
\end{figure}

In Fig. \ref{fig: WKB} we depict how the real and imaginary parts of the QNM frequency of a massless scalar field vary with respect to the model parameter $\alpha$, for $l=1$,  in the case of the WKB approximation method. As we can see, the effect is more significant in the ansatz2 case. We mention here that although the accuracy  of the WKB approximation method depends on the effective potential of the black hole, practically it is more accurate for $l>n$ modes. 

Finally, for comparison, in Table \ref{tab: result table} we summarize the results for a massless scalar field obtained using the WKB approximation method and the time-domain characteristic integration method. And in Table \ref{tab: result table_EM} we give the results for electromagnetic fields.

\begin{table}[]
    \centering
    \resizebox{1.00\columnwidth}{!}{
    \begin{tabular}{|c|c|cccc|cccc|}
    \hline
    \multirow{2}{*}{$\alpha$} & \multirow{2}{*}{$l$} &
    \multicolumn{4}{c|}{ansatz1}          & \multicolumn{4}{c|}{ansatz2}           \\ \cline{3-10} 
                           &                    & WKB(Re) & WKB(Im) & Int(Re) & Int(Im) & WKB(Re) & WKB(Im) & Int(Re) & Int(Im) \\ \hline
    \multirow{3}{*}{0.00}  & 0                  & 0.221   & -0.210  & 0.196   & -0.102  & 0.221   & -0.210  & 0.196   & -0.102  \\ \cline{2-10} 
                           & 1                  & 0.587   & -0.195  & 0.559   & -0.209  & 0.586   & -0.195  & 0.559   & -0.209  \\ \cline{2-10} 
                           & 2                  & 0.967   & -0.194  & 0.964   & -0.192  & 0.967   & -0.194  & 0.964   & -0.192  \\ \hline
    \multirow{3}{*}{0.02}  & 0                  & 0.352   & -0.177  & 0.189   & -0.106  & 0.258   & -0.201  & 0.200   & -0.102  \\ \cline{2-10} 
                           & 1                  & 0.591   & -0.196  & 0.562   & -0.216  & 0.590   & -0.196  & 0.572   & -0.213  \\ \cline{2-10} 
                           & 2                  & 0.967   & -0.196  & 0.977   & -0.205  & 0.967   & -0.196  & 0.973   & -0.197  \\ \hline
    \multirow{3}{*}{0.05}  & 0                  & 0.215   & -0.213  & 0.190   & -0.095  & 0.162   & -0.213  & 0.200   & -0.108  \\ \cline{2-10} 
                           & 1                  & 0.610   & -0.204  & 0.508   & -0.234  & 0.600   & -0.195  & 0.572   & -0.227  \\ \cline{2-10} 
                           & 2                  & 0.967   & -0.201  & 0.982   & -0.214  & 0.968   & -0.199  & 0.969   & -0.213  \\ \hline
    \multirow{3}{*}{0.10}  & 0                  & 0.197   & -0.231  & 0.197   & -0.080  & 0.185   & -0.230  & 0.201   & -0.115  \\ \cline{2-10} 
                           & 1                  & 0.560   & -0.218  & 0.496   & -0.182  & 0.578   & -0.250  & 0.558   & -0.237  \\ \cline{2-10} 
                           & 2                  & 0.964   & -0.215  & 0.912   & -0.278  & 0.964   & -0.215  & 0.963   & -0.227  \\ \hline
    \multirow{3}{*}{0.25}  & 0                  & 0.261   & -0.216  & 0.210   & -0.065  & 0.166   & -0.233  & 0.200   & -0.128  \\ \cline{2-10} 
                           & 1                  & 0.540   & -0.212  & 0.494   & -0.152  & 0.543   & -0.250  & 0.545   & -0.249  \\ \cline{2-10} 
                           & 2                  & 0.931   & -0.219  & 0.886   & -0.186  & 0.942   & -0.249  & 0.955   & -0.246  \\ \hline
    \multirow{3}{*}{0.50}  & 0                  & 0.179   & -0.184  & 0.221   & -0.063  & 0.159   & -0.250  & 0.191   & -0.150  \\ \cline{2-10} 
                           & 1                  & 0.546   & -0.234  & 0.453   & -0.170  & 0.525   & -0.260  & 0.436   & -0.258  \\ \cline{2-10} 
                           & 2                  & 0.904   & -0.217  & 0.904   & -0.283  & 0.908   & -0.262  & 0.921   & -0.262  \\ \hline
    \multirow{3}{*}{1.00}  & 0                  & 0.181   & -0.166  & 0.225   & -0.060  & 0.136   & -0.226  & 0.186   & -0.171  \\ \cline{2-10} 
                           & 1                  & 0.513   & -0.187  & 0.424   & -0.123  & 0.494   & -0.264  & 0.442   & -0.222  \\ \cline{2-10} 
                           & 2                  & 0.876   & -0.216  & 0.940   & -0.239  & 0.867   & -0.272  & 0.879   & -0.271  \\ \hline
    \multirow{3}{*}{2.00}  & 0                  & 0.233   & -0.137  & 0.235   & -0.058  & 0.114   & -0.220  & 0.174   & -0.192  \\ \cline{2-10} 
                           & 1                  & 0.497   & -0.186  & 0.443   & -0.102  & 0.460   & -0.263  & 0.413   & -0.237  \\ \cline{2-10} 
                           & 2                  & 0.843   & -0.209  & 0.910   & -0.227  & 0.820   & -0.277  & 0.839   & -0.274  \\ \hline
    \end{tabular}
    }
    \caption{Values of the quasinormal frequencies for a massless scalar field propagating in the $f(T)$ black hole, for both ansatz1 and ansatz2, based on the WKB method and the characteristic integration method. The $f(T)$ model parameter $\alpha$ varies from 0.02 to 2.00 in units of $2M=1$, while the multipole parameter $l$  acquires the values  0, 1 and 2. }
    \label{tab: result table}
\end{table}

\begin{table}[]
    \centering
    \resizebox{1.00\columnwidth}{!}{
    \begin{tabular}{|c|c|cccc|cccc|}
    \hline
    \multirow{2}{*}{$\alpha$} & \multirow{2}{*}{$l$} &
    \multicolumn{4}{c|}{ansatz1}          & \multicolumn{4}{c|}{ansatz2}           \\ \cline{3-10} 
                           &                    & WKB(Re) & WKB(Im) & Int(Re) & Int(Im) & WKB(Re) & WKB(Im) & Int(Re) & Int(Im) \\ \hline
    \multirow{3}{*}{0.00}  & 1                  & 0.497   & -0.185  & 0.499   & -0.188  & 0.497   & -0.185  & 0.499   & -0.188  \\ \cline{2-10} 
                           & 2                  & 0.915   & -0.190  & 0.919   & -0.190  & 0.915   & -0.190  & 0.919   & -0.190  \\ \cline{2-10} 
                           & 3                  & 1.314   & -0.191  & 1.314   & -0.191  & 1.314   & -0.191  & 1.314   & -0.191  \\ \hline
    \multirow{3}{*}{0.02}  & 1                  & 0.499   & -0.190  & 0.498   & -0.199  & 0.498   & -0.188  & 0.496   & -0.205  \\ \cline{2-10} 
                           & 2                  & 0.914   & -0.194  & 0.919   & -0.203  & 0.914   & -0.193  & 0.928   & -0.195  \\ \cline{2-10} 
                           & 3                  & 1.311   & -0.194  & 1.316   & -0.193  & 1.311   & -0.194  & 1.315   & -0.192  \\ \hline
    \multirow{3}{*}{0.05}  & 1                  & 0.503   & -0.200  & 0.411   & -0.185  & 0.500   & -0.193  & 0.490   & -0.200  \\ \cline{2-10} 
                           & 2                  & 0.912   & -0.200  & 0.908   & -0.210  & 0.913   & -0.198  & 0.920   & -0.210  \\ \cline{2-10} 
                           & 3                  & 1.307   & -0.199  & 1.316   & -0.204  & 1.307   & -0.199  & 1.313   & -0.203  \\ \hline
    \multirow{3}{*}{0.10}  & 1                  & 0.511   & -0.220  & 0.412   & -0.159  & 0.491   & -0.209  & 0.471   & -0.224  \\ \cline{2-10} 
                           & 2                  & 0.906   & -0.209  & 0.899   & -0.209  & 0.905   & -0.210  & 0.908   & -0.225  \\ \cline{2-10} 
                           & 3                  & 1.299   & -0.206  & 1.318   & -0.243  & 1.296   & -0.211  & 1.305   & -0.218  \\ \hline
    \multirow{3}{*}{0.25}  & 1                  & 0.588   & -0.343  & 0.388   & -0.146  & 0.439   & -0.241  & 0.442   & -0.227  \\ \cline{2-10} 
                           & 2                  & 0.884   & -0.233  & 0.850   & -0.171  & 0.882   & -0.243  & 0.880   & -0.243  \\ \cline{2-10} 
                           & 3                  & 1.274   & -0.219  & 1.177   & -0.198  & 1.277   & -0.238  & 1.271   & -0.239  \\ \hline
    \multirow{3}{*}{0.50}  & 1                  & 0.421   & -0.173  & 0.393   & -0.144  & 0.418   & -0.230  & 0.430   & -0.233  \\ \cline{2-10} 
                           & 2                  & 0.840   & -0.208  & 0.878   & -0.269  & 0.844   & -0.253  & 0.844   & -0.253  \\ \cline{2-10} 
                           & 3                  & 1.234   & -0.215  & 1.133   & -0.120  & 1.237   & -0.255  & 1.240   & -0.254  \\ \hline
    \multirow{3}{*}{1.00}  & 1                  & 0.409   & -0.154  & 0.393   & -0.141  & 0.390   & -0.227  & 0.385   & -0.223  \\ \cline{2-10} 
                           & 2                  & 0.804   & -0.197  & 0.806   & -0.219  & 0.803   & -0.260  & 0.805   & -0.261  \\ \cline{2-10} 
                           & 3                  & 1.190   & -0.212  & 1.158   & -0.094  & 1.187   & -0.266  & 1.190   & -0.265  \\ \hline
    \multirow{3}{*}{2.00}  & 1                  & 0.395   & -0.145  & 0.391   & -0.129  & 0.359   & -0.222  & 0.353   & -0.254  \\ \cline{2-10} 
                           & 2                  & 0.770   & -0.185  & 0.797   & -0.193  & 0.756   & -0.262  & 0.762   & -0.262  \\ \cline{2-10} 
                           & 3                  & 1.140   & -0.204  & 1.168   & -0.074  & 1.126   & -0.272  & 1.125   & -0.272  \\ \hline
    \end{tabular}
    }
    \caption{Values of the quasinormal frequencies for an electromagnetic field propagating in the $f(T)$ black hole, for both ansatz1 and ansatz2, based on the WKB method and the characteristic integration method. The $f(T)$ model parameter $\alpha$ varies from 0.02 to 2.00 in units of $2M=1$, while the multipole parameter $l$  acquires the values  0, 1 and 2. }
    \label{tab: result table_EM}
\end{table}

\section{Conclusions and discussion}
\label{sec: conclusion}

In this work we have calculated the Quasinormal modes frequencies of a test massless scalar field around   static black holes solutions in $f(T)$ gravity. Focusing on quadratic $f(T)$ modifications, which is a good approximation for every realistic $f(T)$ theory, we first extracted the spherically symmetric solutions using the perturbative method, imposing two ans$\ddot{\text{a}}$tzes for the metric functions, which suitably quantify the deviation from the  Schwarzschild solution. In both cases the deviation is larger near the event horizon, since the $f(T)$ modification leads to larger slope as well as to larger radial coordinate for the1 event horizon. 

As a next step we extracted the effective potential,  and we showed that the deviation from General Relativity mainly happens for negative values of the tortoise coordinate $r^*$. In particular, for ansatz1 the effective potential acquires a  non-zero asymptotic value, which results from the fact that the two metric functions $N(r)$ and $F(r)$ have different roots. On the other hand, for ansatz2 the effective potential exhibits vanishing asymptotic value, since $N(r)$ and $F(r)$  have the same root. 
 
Finally, we calculated the QNM frequencies of the obtained solutions. Firstly, we solved the Schr$\ddot{\text{o}}$dinger-like equation numerically using the discretization method, and then we extracted the frequency and the time evolution  of the dominant mode applying the function fit method. Secondly, we  performed a semi-analytical calculation by applying the WKB method with the $P_6^6$ Pade approximation. The main results were presented in various figures, and  were summarized in Table \ref{tab: result table} and Table \ref{tab: result table_EM}. As we showed, the results for $f(T)$ gravity are different comparing to General Relativity, and in particular we obtain a different slope and period of the field decay behavior for different values of the model parameter $\alpha$.

Concerning the accuracy of our methods, we can see that the results of the integration and WKB methods agree, however small differences appear due to the limitations of the approximation and numerical accuracy of the two methods. On one hand, the error can arise from the order choice of the WKB approximation. In principle, one can further develop the WKB method to higher orders and agree better with accurate numerical result. On the other hand, for the cases where the quasinormal ringing is very short, the extraction of frequencies with the characteristic integration method can not give satisfactory accuracy. Since in the present study we focus on the first estimate on the QNMs as well as their dynamics within f(T), we leave the higher order calculation as well as gravitational perturbations for future work.

In summary, we have shown that $f(T)$ gravity does have an effect on  the QNM frequency, comparing to General Relativity. This result may be important under the light of the increasing accuracy in gravitational-wave observations from binary systems, in which the characteristic oscillation modes can provide interesting information. Hence, the whole analysis could be used as an additional tool in order to test General Relativity and examine whether torsional gravitational modifications are possible.

\acknowledgments

We are grateful to Geyu Mo, Shengfeng Yan, Zhao Li, and Rui Niu for helpful discussions.
This work is supported in part by the National Key R\&D Program of China (2021YFC2203100), by the NSFC (11961131007, 11653002), by the Fundamental Research Funds for Central Universities, by the CSC Innovation Talent Funds, by the CAS project for young scientists in basic research (YSBR-006), by the USTC Fellowship for International Cooperation, and by the USTC Research Funds of the Double First-Class Initiative. ENS acknowledges participation in the COST Association Action CA18108 ``{\it Quantum Gravity Phenomenology in the Multimessenger Approach (QG-MM)}''. All numerics were operated on the computer clusters {\it LINDA} \& {\it JUDY} in the particle cosmology group at USTC.


\bibliographystyle{JHEP}
\bibliography{QNM}

\end{document}